# Hybridization effects in CeCoIn$_5$ observed by angle-resolved photoemission

A. Koitzsch[1,2], S. V. Borisenko[1], D. Inosov[1], J. Geck[1], V. B. Zabolotnyy[1], H. Shiozawa[1], M. Knupfer[1], J. Fink[1,4], B. Büchner[1], E. D. Bauer[3], J. L. Sarrao[3] and R. Follath[4]

[1]*IFW Dresden, P.O. Box 270116, D-01171 Dresden, Germany*

[2] *Institut für Festkörperphysik, Technische Universität Dresden, D-01062 Dresden, Germany*

[3] *Los Alamos National Laboratory, Los Alamos, New Mexico 87545, USA*

[4] *BESSY GmbH, Albert-Einstein-Strasse 15, 12489 Berlin, Germany*

**Abstract**: We have investigated the low-energy electronic structure of the heavy fermion superconductor CeCoIn$_5$ by angle-resolved photoemission. We focus on the dispersion and the peak width of the prominent quasi-two-dimensional Fermi surface sheet at the corner of the Brillouin zone as a function of temperature along certain k-directions with a photon energy of hν = 100 eV. We find slight changes of the Fermi vector and an anomalous broadening of the peak width when the Fermi energy is approached. Additionally we performed resonant ARPES experiments with hν = 121 eV. A flat f-derived band is observed with a distinct temperature dependence and a k-dependent spectral weight. These results, including both off- and on-resonant measurements, fit qualitatively to a two level mixing model derived from the Periodic Anderson Model.

# 1. Introduction

The various unusual properties of rare earth based heavy fermion (HF) materials arise from the subtle interplay between the f-electrons and the conduction electrons. The f-electron orbital is well localized at the core allowing in most cases only a negligible direct f-f overlap. This picture has been the starting point of the single impurity Anderson model where the rare earth atom (e.g. Ce) is viewed as an impurity within a sea of metallic states, without explicitly considering the translational symmetry of the solid. To rationalize the large effective masses of the charge carriers it is assumed that the f-electrons interact with the conduction electrons, thus renormalizing the electronic structure in the vicinity of the Fermi Energy ($E_F$). The degree of the overlap between f-states and the conduction bands, or the degree of hybridization, decisively influences the properties of the material. At low enough temperatures it facilitates the screening of the large magnetic f-moment by the conduction electrons. This picture is conceptually appealing: the light and mobile conduction electrons become heavy by being locked to the immobile magnetic f-states.

The fraction of the f-electrons that hybridize with the conduction electrons and become itinerant is normally small while the rest remains localized. This dual nature of the f-electrons is well established and can be directly monitored by photoemission spectroscopy. The f-derived spectral weight appears at $E_F$ for the hybridized part while the localized part resides about 2 eV below $E_F$. With the recent progress in resolution and measuring efficiency the detailed angle-resolved investigation of the Kondo-like $E_F$ peak became possible. Angle-resolved photoemission spectroscopy (ARPES) is the method of choice to detect possible dispersions of this peak and k-dependent hybridization. This information cannot be achieved by transport and thermodynamic measurements as they integrate over k. In fact much effort has been undertaken in this direction. For Ce-based compounds various k-dependent effects have been reported underlining the importance of the lattice [1, 2, 3, 4]. Recently the direct observation of a f-derived quasiparticle band dispersion was reported in CeIrIn$_5$ [5]. Although the f-electrons behave dominantly localized in this compound, a small but finite fraction hybridizes and determines thereby the low energy properties.

The CeTIn$_5$ (T = Co, Rh, Ir) family has attracted much attention because it shows a number of remarkable properties such as unconventional superconductivity and quantum criticality [6, 7]. In particular CeCoIn$_5$ has a superconducting transition temperature T$_c$ = 2.3 K which is unusually high for heavy fermion superconductors. CeRhIn$_5$ is an antiferromagnet with T$_N$ = 3.8 K [8]. The non-Fermi liquid behavior accompanying the quantum critical region of the magnetic field-temperature phase diagram of CeCoIn$_5$ shows signatures of antiferromagnetic fluctuations [9, 10] which might also be involved in the mechanism of the pairing. The crystallographic and the electronic structure of CeTIn$_5$ show two-dimensional aspects. The crystal is built by CeIn$_3$ and TIn$_2$ layers stacked along the c-axis. The electronic structure shows a prominent quasi-cylindrical sheet at the corner of the Brillouin zone. This has been observed by de Haas van Alphen experiments [11, 12] in combination with band structure calculations [13, 14, 15]. The two-dimensionality increases in going from Ir to Rh and Co [11, 12].

Here we investigate CeCoIn$_5$ by ARPES concentrating on the vicinity of the Fermi energy to monitor the hybridization of the f-states with the conduction electron states. A resonance measurement is performed to enhance the visibility of the f-electrons. We find characteristic changes of the spectral response with temperature for both, off and on resonant measurements. The results are in qualitative agreement with a simplified two-level mixing model which has its roots in the Periodic Anderson model (PAM).

## 2. Experiments

The ARPES experiments were carried out using radiation from the U125/1-PGM beam line and an angle multiplexing photoemission spectrometer (SCIENTA SES 100) at the BESSY synchrotron radiation facility. The x-ray photoemission (XPS) measurement was performed using a SCIENTA SES 200 based spectrometer and Al Kα excitation. We present photoemission data taken with hν = 90, 100, 121 and 1486.6 eV with an energy resolution of 70 meV for the Fermi surface (FS) map in Fig. 1 and 40 meV for the spectra in Fig. 2 and 3 except for the XPS spectrum with an energy resolution of 0.5 eV. The angu-lar resolution was 0.2°. Single crystals of CeCoIn$_5$ were grown in In flux. For the measurements the crystals were cleaved *in situ* at room temperature.

## 3. Results and discussion

Figure 1 shows a Fermi surface (FS) map measured at T = 25 K with photon energy hν = 100 eV and a schematics of the Brillouin zone (BZ). The cleaved surface is parallel to the Γ-M-X plane. The most important feature of this map is the closed structure at the corner of the BZ with the shape of a rounded square. It consists actually of two components, α and β, as will be shown below. Here α is part of the 135 band in the notation scheme of Ref. [15] and β belongs to the 133 band. The α sheet represents the prominent quasi-cylindrical part of the FS. From a comparison of photoemission data taken along the high symmetry directions of the BZ to band structure calculations we conclude that with a photon energy of hν = 100 eV a FS cut in the vicinity of the Z-plane is probed. For a detailed discussion of the band structure see [16].

To investigate in detail the low energy physics arising from the two dimensional FS sheet α we collected energy distribution maps (EDM's) along the light blue bars in Fig. 1. One of the bars corresponds to the zone diagonal. The other bar is, for reasons explained below, somewhat off the AR high symmetry direction.

Fig. 2 presents EDMs along the zone diagonal at 130 and 25 K. The EDM shows two bright bands. One band centered at k = 0.4 and -0.4 eV is only partly visible. It belongs to the 131 band which also crosses the Fermi energy. Here we concentrate on the electron like parabola shaped band with the minimum well below E = 0.4 eV which forms the quasi-cylindrical FS sheet centered at the BZ corner. According to the FS map there should be one more band crossing the Fermi energy stemming from the bright short arc next to the quasi-cylindrical band at somewhat lower k. However, this structure is absent in the EDM. It varies greatly in intensity from cleavage to cleavage and is probably surface related [16]. To study the intrinsic quasi-cylindrical sheet we focus on the dataset in Fig. 2 where its intensity is suppressed. Careful comparison between the two different temperatures reveals small changes of this band in the vicinity of the Fermi energy. A slight bending of the band seems to be present as marked with a white arrow. To substantiate this observation and avoid deception due to the false color representation we refer directly to the data in Fig. 2c. Here we show momentum distribution curves (MDC), i.e. horizontal cuts trough the images in panel a) and b). The energy positions of the MDCs at 130 K and 25 K are marked by the roman numbers. At energies away from the Fermi energy the MDCs for different temperatures coincide over several hundreds of meV. Only the MDC comparison at $E_F$, i.e. for the top 130 K/25 K MDC pair, shows slight deviations: the blue MDC is shifted towards lower momenta. The upturn for the lowest MDCs at the right hand side is due to the 131 band. There seems to be another weak feature in between this 131 band and the parabola shaped band under consideration. Its approximate position is marked with a black dot in Fig. 2c. It has the shape of a very small peak at high energies and develops into a shoulder near the Fermi energy. This feature is identical to the β sheet and its presence is consistent with band structure calculations [16]. In the upper part of panel 2c we present a comparison of the MDCs at T = 25 K from the Fermi energy (E = 0 meV) and at higher energies (E = 147 meV). The conventional behavior would be an increase of the MDC width for higher energies. This is due to the fact that the lifetime of the excitation, i.e. its scattering length, decreases for higher binding energies. However, here we observe the opposite effect. The MDC at higher energy is clearly narrower than at low energy. We speculate that at very low energies an additional scattering process is present involving f-states.

It may be argued that two features present in the data may have some influence on the overall line shape and introduce some uncertainty about the bending of the dispersion and the broadening. To avoid those complications, we present in Fig. 3 data offset from the AR direction because here the splitting of α and β bands is larger than exactly on the high symmetry line. Clearly, two almost parallel bands of similar magnitude are observed as is shown at 130 K (Fig 3a) and 25 K (Fig. 3b). They obey a parabola-like shape with a band minimum below 0.5 eV. The inner band, which is more intense, corresponds to the α sheet of the 135 band whereas the outer band corresponds to the β sheet of the 133 band. Comparing the two temperatures we find overall agreement except for the vicinity of the Fermi energy. A similar bending to that discussed in Fig. 2 seems to be present for the inner α sheet but is hardly visible for the outer β sheet. In Fig. 3c we present MDCs from Figs. 3a and 3b to check this visual impression. At high energies, away from the Fermi energy, the MDCs at 25 K and 130 K are nearly identical. A systematic deviation is present only for the MDC at $E_F$: The inner peak is shifted somewhat towards lower momenta for low temperatures while for the outer peak such a shift is not recognizable but cannot be ruled out due to the signal to noise ratio. Comparison between an MDC at $E_F$ to an MDC at E = 180 meV in the upper part of Fig. 3c shows that the MDC at $E_F$ is broader than the one at E = 180 meV, which is consistent with the observation for the ZA direction. To explain the above observations, we assume that at low energy and at low temperature a weak hybridization occurs between the valence band states and the f-states. This will shift $k_F$ somewhat to adjust the FS volume to the number of available charge carriers and open an additional scattering channel for low energy excitations. In this picture the hybridization seems to be stronger for the α sheet than for the β sheet for this particular k-space location and for the given temperature.

In order to further explore the role of the f-states we present resonant photoemission data. The cross section for the photoemission process favors emission from the Co 3d orbitals for hν = 100 eV. For hν = 121 eV the Ce 4d –4f threshold is reached and the emission from the Ce 4f orbitals is significantly enhanced. Figure 4 shows a comparison of the valence band (VB) spectrum off the Ce 4d resonance and on resonance. Off resonance, the VB is dominated by the Co 3d derived peak at E = -0.7 eV. On resonance, the 4f states are enhanced and an intense peak at E = -1.4 eV appears which corresponds to the $4f^0$ final state. A smaller enhancement is found at the Fermi energy, where the itinerant 4f states and their spin orbit split sideband are expected. The small relative magnitude of the near $E_F$ f-weight indicates weak hybridization. Thus, elements of both localized and itinerant behavior are found in CeCoIn$_5$. Presumably, it is the small but non-vanishing itinerant f-electron fraction that determines the low-energy physics and, hence, the low-temperature properties of the system.

Comparing the energy position of the $f^0$ peak E = -1.4 eV with CeIrIn$_5$ (E = -2.5 eV) and CeRhIn$_5$ (E = -2.5 eV) a large difference is found. This seems surprising with respect to the isoelectronic character of Co, Rh and Ir. To exclude that the $f^0$ peak at –1.4 eV indicates an anomalous surface contribution, unrelated to the bulk properties, an additional XPS (hν =1486.6 eV) valence band spectrum is presented in Fig. 4. At this photon energy the electron escape length is increased by a factor of five compared to hν = 121 eV and the cross section ratio between Ce 4f and Co 3d is 0.6. As can be seen (Fig. 4) the $f^0$ peak appears approximately at the same position, ruling out dramatic differences of the electronic structure between surface and bulk at this stage. The difference to CeRhIn$_5$ and CeIrIn$_5$ is puzzling. It could be related to properties of the electronic structure such as the valence bandwidth and the hybridization strength or the appearance of different cleavage surfaces among the materials. Additional bulk sensitive measurements are needed to clarify this point. Anyway, the 4f electrons tend to be much more localized at the surface than in the bulk [17], therefore, we assume that the small itinerant 4f fraction in our spectra originates from bulk states, although 4d-4f resonance photoemission is a surface sensitive technique.

In Fig. 5 we study the angle dependence of the near $E_F$ f-weight. Fig. 5a shows the off-resonant energy dependence of the spectral weight along the horizontal blue line in Fig.1 at hν = 90 eV. The α band characterizing the cylindrical FS-sheet is clearly observed. The β sheet is very weak probably due to matrix element effects. Fig 5b shows the integrated energy distribution curves (EDC) from inside the dashed k-intervals. No f-related peaks are found in the vicinity of $E_F$. Now we compare the off resonant measurement in Fig 5a to the on resonant measurement taken with hν = 121 eV presented in Fig 5c. Again the electron like α band, which consists mainly of Co 3d and In 5p states, is detected but not the β sheet. Additionally, we find an intense flat band at $E_F$ and a weak sideband at E = -0.3 eV. This flat band with significant f-character is more intense "outside" the parabolic band (i.e. at the hole-like side) than "inside" (i.e. at the electron-like side). There is also featureless enhanced intensity in the lower right corner of Fig 5c. This extra intensity could arise from the valence band which is enhanced either purely by matrix element effects, or by the resonance condition itself: note that the $f^0$ peak is located at E = -1.4 eV and rather broad (see Fig. 2). In the latter case the k-dependence of the spectral $f^0$ weight would resemble that of the $f^1$ component.

The EDC spectra in panel Fig. 5d show a resolution limited peak "outside" (lower panel), i.e. the hole-like side, of the electron band and its spin orbit split component at E = -0.3 eV. "Inside" the electron parabola the peak is much weaker (upper panel). Increasing the temperature decreases the intensity of the peak at the Fermi energy as shown in Figs. 5e-

h. At T = 105 K the peak is still visible in the EDC spectrum (Fig. 5f) but considerably broadened, while the flat band is hardly discernible anymore (Fig. 5e). At 180 K the peak still exists but there is a less dramatic change than between 20 K and 105 K. No feature at $E_F$ is present in the upper panel of Fig. 5h, i.e. at the electron like side; thus, our data directly reveals the evolution of the heavy mass quasiparticle band with temperature. Additionally we find a pronounced k-dependence of the spectral weight distribution. The f-derived band is much stronger outside the parabola shaped band than inside (Fig 5d). There must be also at least one additional band crossing (from β) outside the α sheet, which is not discernible in the 121 eV data. This crossing may contribute to the imbalance of the spectral weight distribution away from $k_F$.

A schematic of the band structure is displayed in Fig. 6 and may help explain the imbalance of the spectral weight distribution. The overall dispersion of the non-f valence bands is well described by band structure calculations [16]. For the theoretical description of the distribution of the spectral f-weight shown in Fig. 5c we approach the electronic structure from a different point of view, which is based upon the Periodic Anderson model [18, 19, 20]. We assume a normal non f-band which cuts $E_F$ with a sizable slope (see Fig. 6a). The f-sites are modeled by a dispersionless state at a renormalized energy $\varepsilon_f$' above $E_F$. A finite hybridization opens a gap and results in two branches $E_1$ and $E_2$. The Fermi vector is renormalized from $k_F$ to $k'_F$ to accommodate the additional f-electrons and the Fermi velocity is reduced, leading to an increase of the effective mass. The hybridization gap pushes spectral weight towards $E_F$ "outside" the normal band and away from $E_F$ "inside" the normal band. Fig 6b shows the spectral weight expected in an off-resonant photoemission experiment. It is dominated by the conventional non-f band except for small renormalization near $E_F$. On the other hand, for the on resonant condition (Fig 6c), the spectrum is dominated by the f-weight outside the conventional band. This description is in qualitative agreement with our experimental observations (Fig 2, 3, 5). Although the off resonant data in Fig. 2 + 3 and the angle-integrated data in Fig 4 show only weak hybridization effects, the distribution of spectral f-weight is inhomogeneous. The k-space location where the f-related spectral weight is observed, namely in between the quasi-cylindrical bands, agrees also with LDA band structure calculations, regardless of the bandwidth and the dispersion.

The way the f-weight is distributed over the FS, i.e. the k-dependence of the heavy quasiparticles, may have significant consequences for superconductivity in CeCoIn5. If antiferromagnetic fluctuations with characteristic momentum **q** are involved into the pairing mechanism, the interaction among two quasiparticles (momenta **k$_1$**, **k$_2$**) may depend on **q** via **q** = **k**$_1$-**k**$_2$, which puts geometrical restrictions on the suitable FS regions and therefore on the distribution of f-weight. It is interesting that the FS of CeCoIn$_5$, and in particular the quasi two dimensional α sheet, resembles the prominent "barrel" FS of the cuprates. Often superconductivity is found in tetragonal crystal structures, which, for unknown reasons, seem beneficial for this phenomenon. One may speculate then *vice versa* that it is the electronic structure of the type found in the high temperature superconductors and CeCoIn$_5$ which promotes superconductivity. The relatively large barrels at the BZ corner occupy a significant k-volume prohibiting this area for features with high density of states. Those are found in between the barrels then, namely as the van Hove singularity for the cuprates and as flat renormalized f bands for Ce*T*In$_5$. There they can be connected by the antiferromagnetic nesting vector **q** = (π/2,π/2). Exactly the latter might represent a precondition for pairing in such systems.

There is evidence from our data (Fig. 5) that the temperature dependence of the heavy quasiparticle band mimics the coherence feature observed in electrical resistivity measurements. We return to the temperature dependence of the heavy fermion band presented in Fig. 5 e-h. Transport measurements [6, 21] found substantial decrease of the resistivity below the coherence temperature T* ≈ 45 K, indicating the formation of a coherent heavy fermion band, while above T* the transport is dominated by scattering from local moments. It has been observed that the phonon scattering is weakly temperature dependent and of negligible magnitude below 100 K [22]. We observe a sharp f-derived band at E$_F$ at 20 K which rapidly broadens when the temperature is increased to 105 K and decreases and broadens more slowly when the temperature is further increased to 180 K. It is suggestive to associate this behavior with the dominant scattering channel found in transport.

## 4. Summary

In summary we investigated the low energy electronic structure of $CeCoIn_5$ by off and on resonant ARPES. Off the Ce 4d resonance at hν = 100 eV, which corresponds to a $k_z$ value in the vicinity of the Z-plane of the tetragonal Brillouin zone, we find small changes of the Fermi vector for the cylindrical α sheet as a function of temperature and an anomalous broadening of the MDC when the Fermi energy is approached along ZA. Similar results are found parallel to the AR direction. We investigated the role of the 4f-electrons directly by resonant ARPES and observe a flat f-derived band close to $E_F$ which shows pronounced k-and temperature dependencies. The k-dependence of the f-states and the small renormalizations found off resonant can be understood within the Periodic Anderson model.

We acknowledge financial support by the DFG SFB 463. Work at Los Alamos was performed under the auspices of the U.S. DOE. We would like to thank S. L. Molodtsov, Yu. Kucherenko, I. Opahle, S. Elgazzar, M. Richter, J. D. Denlinger and J. W. Allen for valuable discussions.

# Captions

**Fig. 1**: Fermi surface map at T = 25 K with hv = 100 eV photon energy. Bright yellow denotes regions of high intensity. The first Brillouin zone (BZ) is highlighted by dashed lines. The Fermi surface consists of closed squarish structures centered at the corner of the BZ and bright complex structures around the BZ center. The light blue bars mark the positions of the measurements shown in Fig. 2 and 3.

**Fig. 2**: Energy distribution maps along the zone diagonal (see the diagonal bar in Fig. 1) for T = 130 K (a) and T = 25 K (b) and hv = 100 eV. The white arrows highlight the slight temperature dependence of the dispersion. (c): Momentum distribution curves (MDC) extracted from (a) and (b). The energy position of the blue/red MDC is marked by arrows. The upper blue/red MDC pair corresponds to the Fermi energy. The peak heights have been normalized to one. The MDCs are integrated over an energy window of 12 meV and of 24 meV directly at the Fermi energy. The black/green MDC pair is a comparison of different energies. The E = 147 meV MDC has been shifted along k to enable comparison of the peak width. The black dot highlights the approximate position of a second weak feature.

**Fig. 3**: Energy distribution maps parallel to the zone boundary (see the horizontal bar in Fig. 1) for T = 130 K (a) and T = 25 K (b) and hv = 100 eV. (c): Momentum distribution curves (MDC) extracted from (a) and (b). The energy position of the blue/red MDC pairs is marked by arrows. The upper blue/red MDC pair corresponds to the Fermi energy. The peak heights have been normalized to one. The MDCs are integrated over an energy window of 12 meV and of 24 meV directly at the Fermi energy. The black/green MDC pair is a comparison of different energies. The E = 180 meV MDC has been shifted along k to enable comparison of the peak width.

**Fig. 4**: Comparison of typical valence band spectra measured off- (117 eV) and on-resonance (121 eV) and with high photon energy (1486.6 eV). The off resonant spectra is dominated by a Co 3d derived peak at E = -0.7 eV. On resonance there is a strong en-

hancement around E = -1.4 eV and a smaller enhancement in the vicinity of $E_F$. The high energy valence band features are broadened but match the low energy features. The spectra have been normalized to the peak maximum.

**Fig. 5**: (Upper row) Energy distribution maps (EDM's) along the horizontal blue line in Fig. 1 measured off and on resonance as a function of temperature. Note the parabola shaped band (red circles) and the appearance of the additional flat band (black circles) near $E_F$ in panel (c). (b), (d), (f), (h) : Energy distribution curves integrated from the regions enclosed by the pink or black dashed lines. (upper row) to the left from $k_F$ ("inside" the parabola); (lower row) to the right from $k_F$ ("outside" the parabola). The dashed red line in the lower panels is a guide to the eye indicating approximately the background contribution derived from panel (b).

**Fig. 6**: (a) Schematic view on the Anderson lattice model. The dashed lines are a dispersing valence band and the dispersionless renormalized f-states above the Fermi energy. For a finite hybridization, a gap opens and two branches with k-dependent orbital weight are formed ($E_1$, $E_2$). The $E_F$ crossing is adjusted to accommodate the additional f-electrons. The Fermi velocity is reduced and the effective mass enhanced. (b) Off resonant spectral weight as seen by photoemission. (c) On resonant spectral weight.

**Fig. 1**

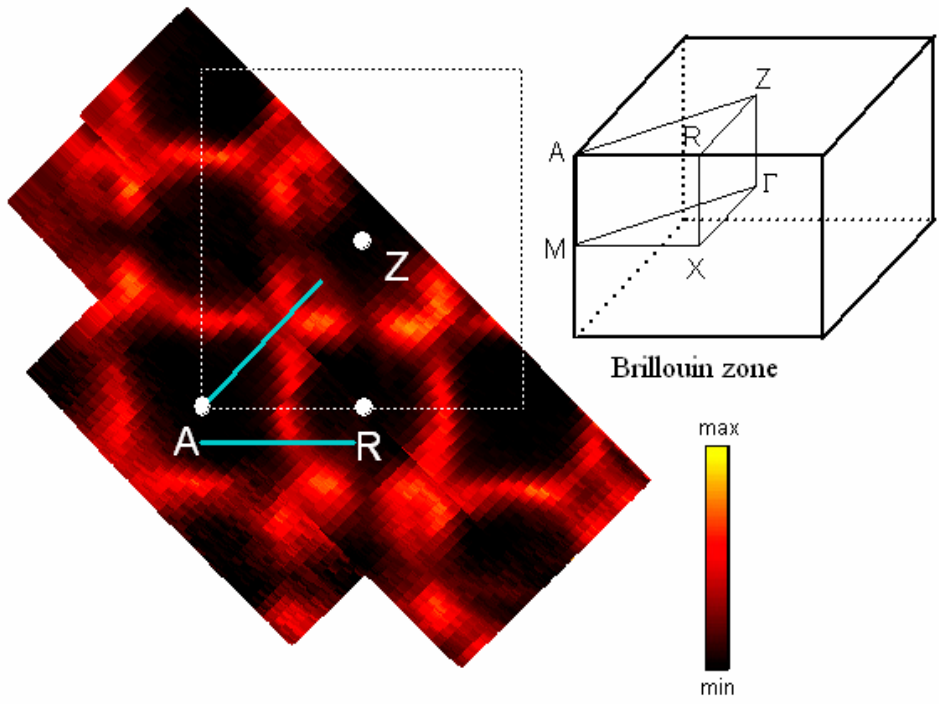

**Fig. 2**

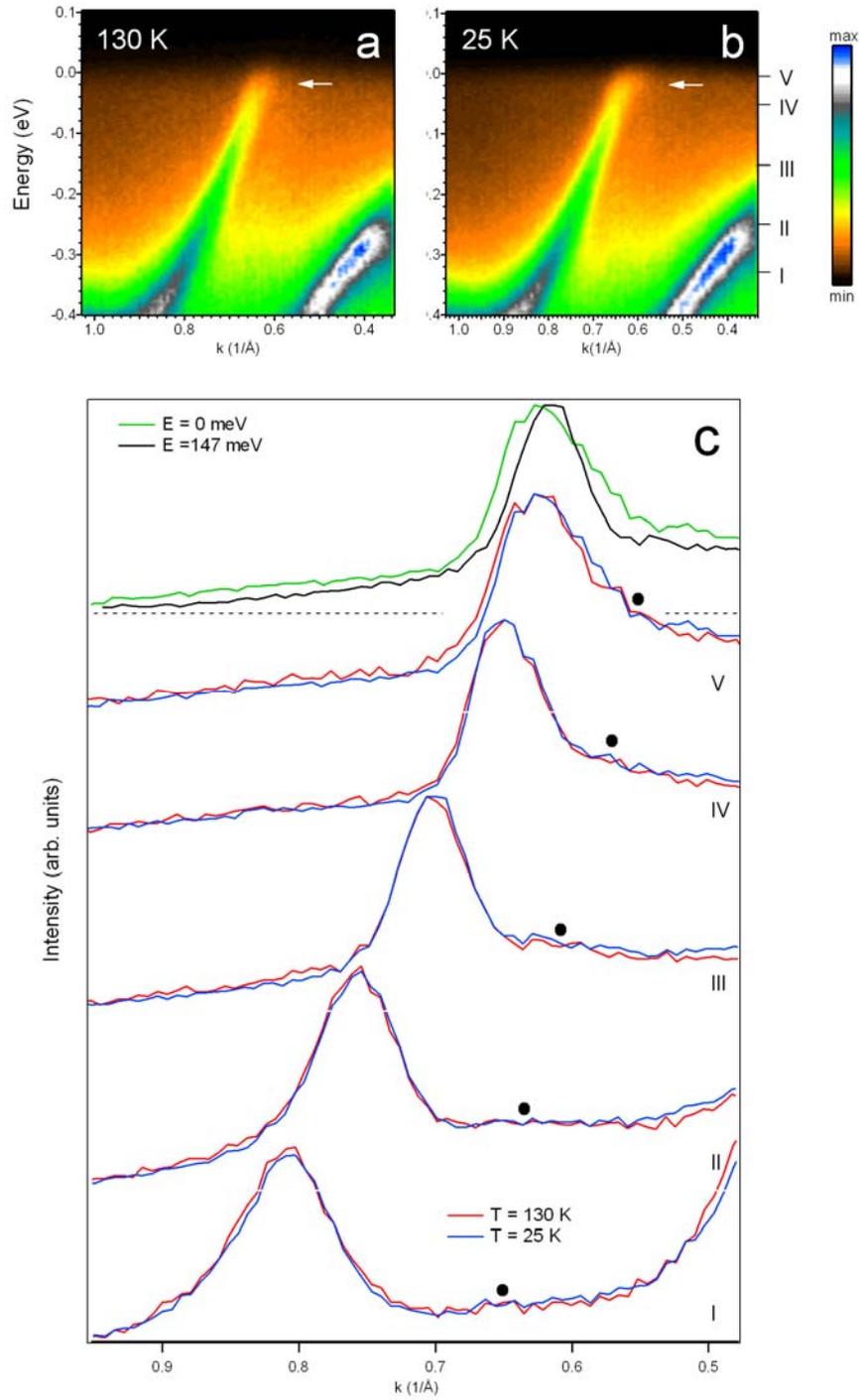

**Fig. 3**

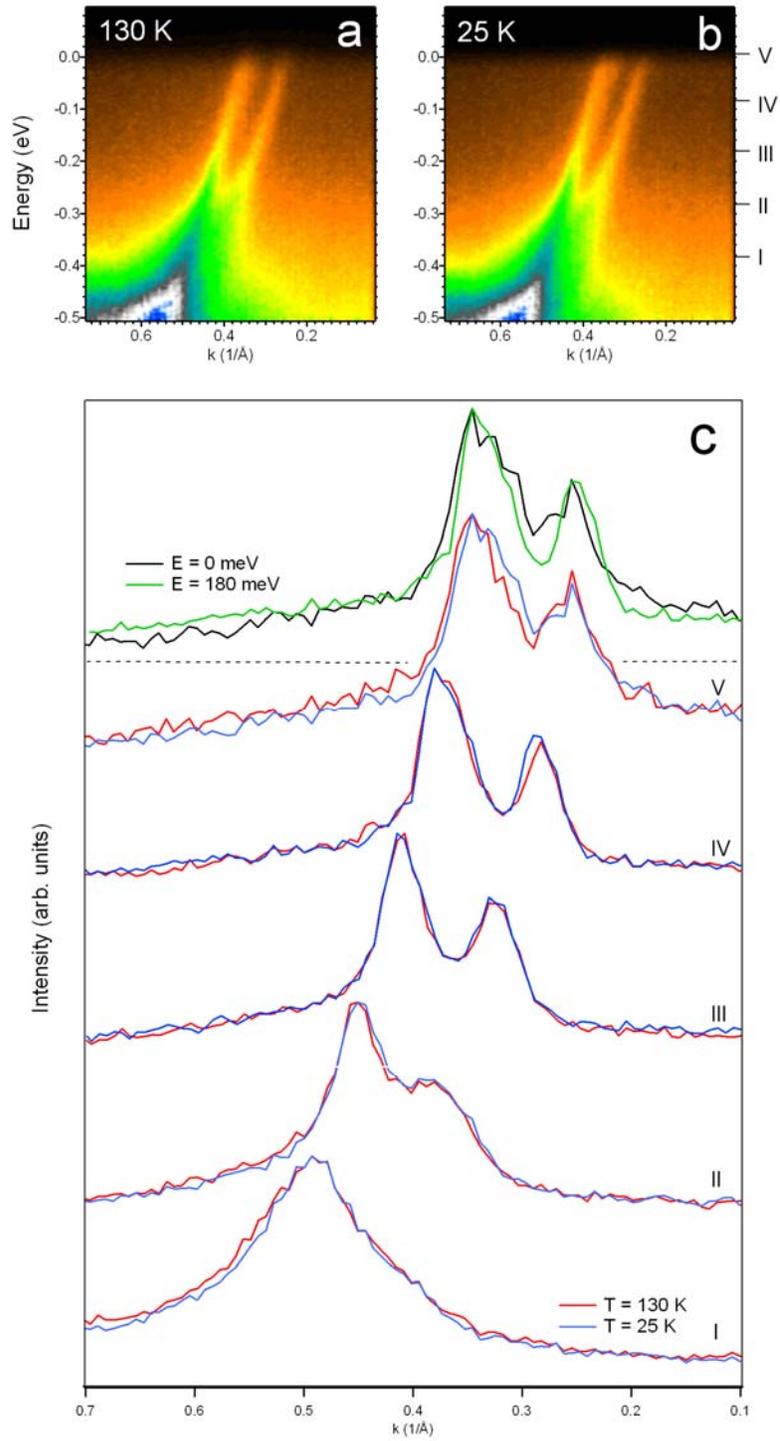

**Fig. 4**

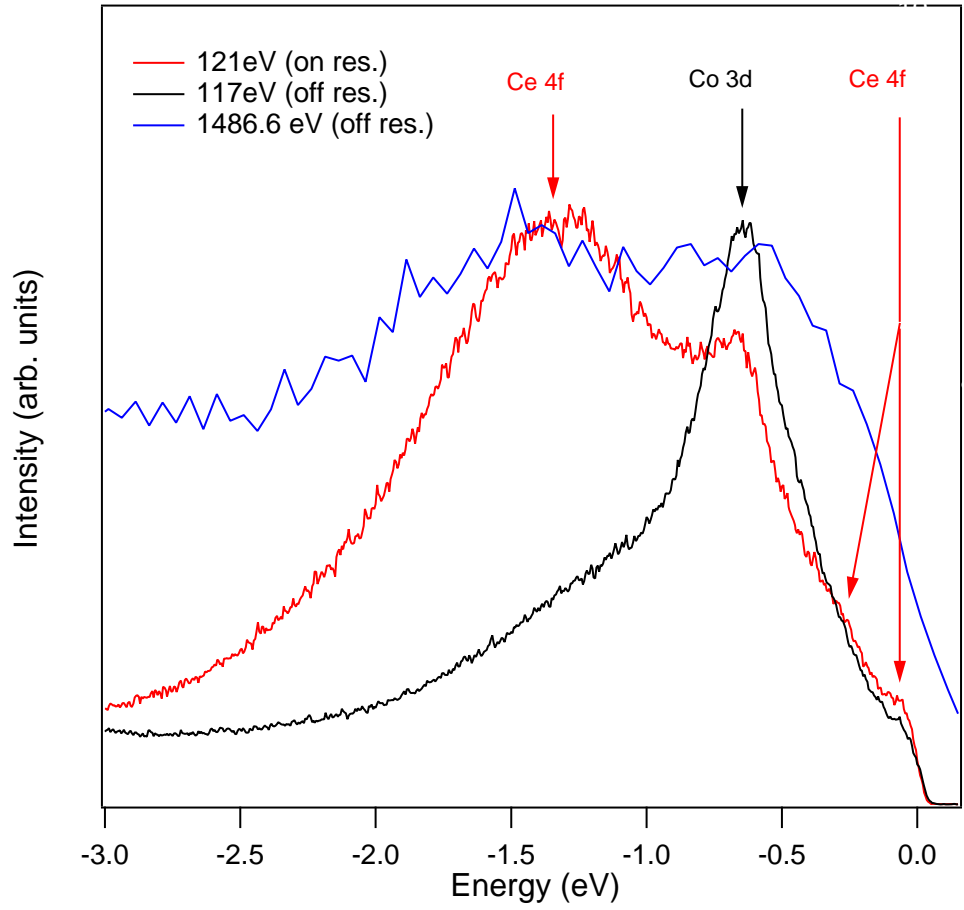

**Fig. 5**

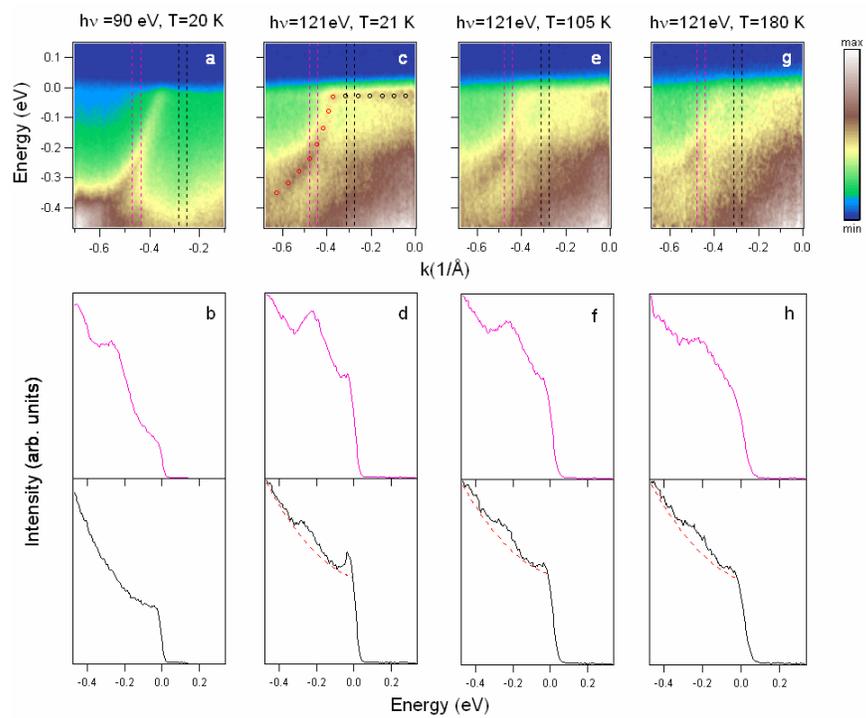

**Fig. 6**

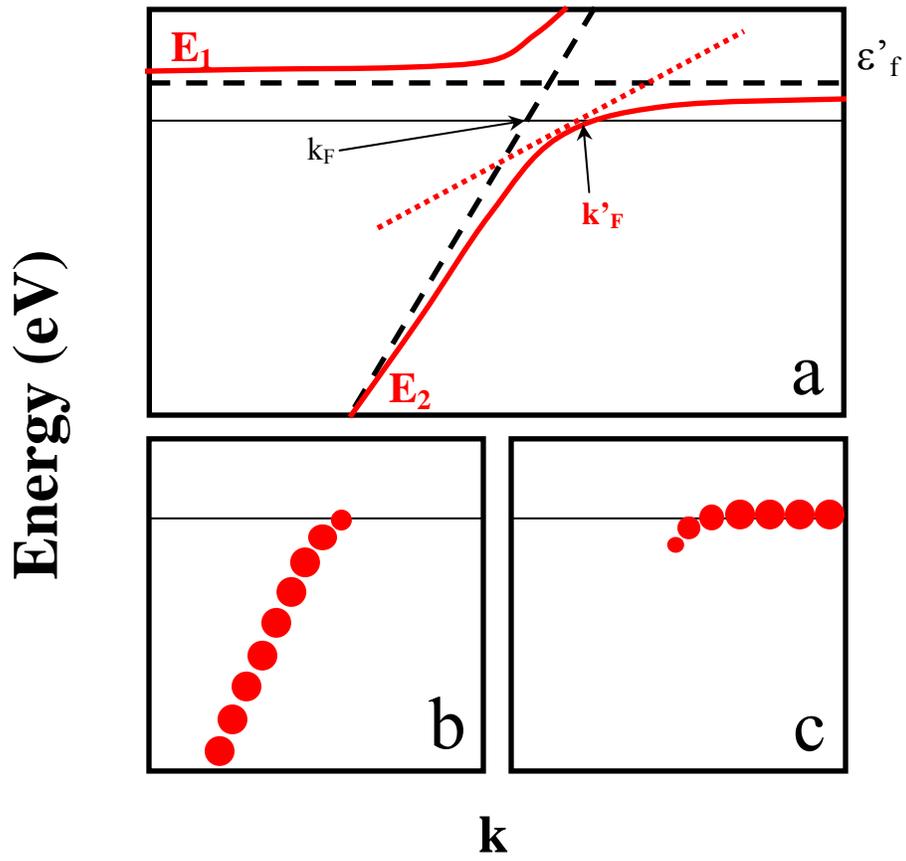